\documentclass{iaus}
\usepackage{times}

\title  {Historical Perspective on Computational Star Formation}

\author                  {Richard B. Larson}

\affiliation {Department of Astronomy, Yale University,\\
                   New Haven, CT 06520-8101, USA\\
                     {richard.larson@yale.edu}}

\pubyear{2010}
\volume{270}
\jname{Computational Star Formation}
\editors{B. G. Elmegreen, J. Alves, J. M. Girart, \& V. Trimble, eds.}
\begin{document}

\maketitle

  The idea that stars are formed by gravity goes back more than 300 years to Newton, and the idea that gravitational instability plays a role goes back more than 100 years to Jeans, but the idea that stars are forming at the present time in the interstellar medium is more recent and did not emerge until the energy source of stars had been identified and it was realized that the most luminous stars have short lifetimes and therefore must have formed recently.  The first suggestion that stars may be forming now in the interstellar medium was credited by contemporary authors to a paper by Spitzer in 1941 in which he talks about the formation of interstellar condensations by radiation pressure, but then oddly says nothing about star formation.  That may be because, as Spitzer later told me, when he first suggested very tentatively in a paper submitted to The Astrophysical Journal that stars might be forming now from interstellar matter, this was considered a radical idea and the referee said it was much too speculative and should be taken out of the paper.  So Spitzer removed the speculation about star formation from the published version of his paper.

  But the idea apparently got around anyway, and it was soon developed further by Whipple in a paper that credited Spitzer for the original suggestion.  Whipple says in a footnote that although his work was first presented in 1942, its publication was delayed by ``various circumstances" until 1946.  By that time, the idea that stars are forming now in the interstellar medium had evidently become respectable enough to be published in The Astrophysical Journal, and Whipple's paper may be the first published presentation of it.  In 1947, Bok \& Reilly called attention to the compact dark clouds in the Milky Way that later became known as Bok globules, and they suggested that these dark globules might be prestellar objects and might form stars, referencing the papers by Spitzer and Whipple.  This suggestion was controversial at the time, and it remained so for many years.  But in 1948 Spitzer, in an article in Physics Today, laid out what are essentially modern ideas about star formation in dark clouds, and he pointed specifically to the dark globule Barnard 68 as a possible prestellar object, or `protostar' as he called it.

  By the 1950s, the theory of star formation had become a popular subject and many papers were written on it.  The most influential one was probably a 1953 paper by Hoyle that introduced the concept of hierarchical fragmentation, whereby a cloud is assumed to collapse nearly uniformly until at some point separating or fragmenting into smaller clouds, which then individually collapse nearly uniformly and repeat the process.  The idea of hierarchical fragmentation remained influential for a long time in theoretical work, even though the assumption of uniform collapse was later disproven by numerical calculations.

  Numerical work on star formation began in a serious way in the 1960s, and I came into the picture in 1965 when the problem of protostellar collapse was suggested to me by my thesis advisor Guido M\"unch at Caltech.  Originally I had grandiose ideas about calculating galaxy formation, but Guido was skeptical and said ``before you try to understand how a galaxy forms, why don't you try to understand how one star forms?"  He also suggested that I talk to Robert Christy, who had recently used numerical techniques to study stellar pulsation, and see if I could use similar techniques to calculate the collapse of an interstellar cloud to form a star.  I thought that this sounded like an interesting and challenging project, and I went to talk to Christy, a nuclear physicist who had worked in the nuclear weapons program at Los Alamos.  He thought that my calculation might be feasible, and he handed me some reprints and preprints, among which was a recently declassified report from the Livermore National Laboratory presenting a numerical method for doing gas dynamics with radiation and shocks that had originally been developed to calculate powerful explosions in the Earth's atmosphere.  I realized that I could use some of the same techniques for the star formation problem, and I also recognized in this report the origin of what became the most widely used method for calculating stellar evolution, the `Henyey method', which had been derived from the same Livermore bomb code by taking out the hydrodynamics.  Many of the numerical techniques later used in astrophysics thus had their origins in nuclear weapons research, perhaps not surprisingly given that a nuclear explosion may be the closest terrestrial counterpart to astrophysics, involving similar physical processes.

  When I began work on the protostellar collapse problem in late 1965, I had no idea what I would find or how far I would get, but I thought that even a start on the problem would be worthwhile.  Along the way I wrote and tested two completely independent codes, Lagrangian and Eulerian, each with its advantages and disadvantages, and I tried as far as possible to replicate my results with both codes to increase my confidence in them.  About a year later in late 1966, I completed my first calculation that had started with something like a Bok globule and ended with a pre-main sequence star.  The basic result was that the collapsing cloud became so centrally condensed that only a tiny fraction of its mass at the center first attained stellar density, becoming a `stellar core' that continued to grow in mass by accretion until eventually acquiring most of the initial cloud mass.  The essential implication of this was that star formation is largely an accretion process.  This was clearly an important result, and I realized that I still had a lot of work to do to demonstrate its correctness and robustness, so I spent another year running more cases and varying the assumptions and approximations involved.  Eventually, after much testing, I acquired considerable confidence in my results, and I presented them in my thesis in 1968.  In my thesis defense I was careful to note that my calculation was still an idealized case assuming spherical symmetry and neglecting rotation and magnetic fields, which seemed unlikely to be realistic.  But one of my examiners, I think it was Peter Goldreich, said ``don't be so apologetic, this is a good calculation and you should publish it."

  Thus encouraged, I published my results in 1969 and presented them at meetings.  They attracted considerable interest, but also received a lot of flak and criticism.  There followed about a decade of debate and controversy over whether my results were correct, with some studies yielding conflicting results and with observers producing apparently conflicting observations showing outflows rather than inflows around newly formed stars.  But Bok was delighted that I had shown how one of his globules could form a star, and he decided to spend his retirement years as a kind of evangelist for Bok globules.  He was vindicated in 1978, when he proudly sent me a photograph he had taken of a dark globule with a Herbig-Haro jet emerging from it, showing that a star had recently formed in this globule.  My vindication came in 1980 when two groups, Winkler \& Newman and Stahler, Shu, \& Taam, published results very similar to mine.  More recently, Masunaga \& Inutsuka in 2000 considerably refined the spherical collapse calculation and again obtained similar results.

  What was learned from all this work that could be credited specifically to the use of numerical methods?  Looking back, I think that the most important result of my work might have been the very first one that I found when I got my first collapse code running at the end of 1965.  I had written a simple Lagrangian code to calculate isothermal collapse, and the first successful run with this code showed the runaway growth of a sharp central peak in density.  I plotted the density distribution logarithmically and noticed that it was approaching a power-law form with $\rho \propto r^{-2}$, a form similar to that of a singular isothermal sphere, even though the cloud was collapsing almost in free fall.  This power-law behavior extended to smaller and smaller radii as the collapse continued.  Although this result was unexpected, I realized that it could be understood qualitatively in terms of the inward propagation of a pressure gradient from the boundary, and I later found an asymptotic similarity solution showing this behavior and was able to show that the numerical solution was evolving toward it, giving me increased confidence in the result.  I also later learned that at about the same time Michael Penston had been doing similar work and finding similar results, and he independently derived the same similarity solution.  This `Larson-Penston solution', as it has been called, has been perhaps the most enduring result of that early work, and similar asymptotic similarity solutions have been found for a variety of other more realistic collapse problems, including non-isothermal and non-spherical collapse and even collapse with rotation and magnetic fields.

  Concerning collapse with rotation, I tried in 1972 to calculate the collapse of a rotating cloud with axial symmetry, but this time I got it wrong.  My numerical resolution in 2 dimensions, limited by the computers then available, turned out to be inadequate to follow the development of a sharp central density peak, and my calculation showed instead the formation of a ring.  Later when we got a bigger computer, I repeated the calculation with a finer grid and got a smaller ring, causing me to wonder whether the ring might go away completely with infinite resolution.  The first person to get it right was Michael Norman, and in 1980 Norman, Wilson, \& Barton showed that when sufficient care is taken to ensure adequate resolution at the center, the result is not a ring but a centrally condensed disk that evolves in a quasi-oscillatory fashion toward a central singularity.  This result was later confirmed in more detail in 1995 by Nakamura, Hanawa, \& Nakano, who also derived an asymptotic similarity solution similar in form to the Larson-Penston solution describing the evolution of the disk toward a central singularity.  Finally in 1997, Basu showed that a similar asymptotic similarity solution describing evolution toward a central singularity can be derived even when a magnetic field is included in addition to rotation and when ambipolar diffusion is properly included in the calculation.

  What these results show is that in all of these cases, star formation begins with the runaway development of a central singularity in the density distribution.  This conclusion now seems to be universal, and even in more realistic 3-dimensional simulations of the formation of systems of stars, the formation of each simulated star or `sink particle' always begins with the sudden appearance of a near-singularity in the density distribution in a place where local collapse is occurring.  This might now seem an unsurprising result because stars are essentially mass points or singularities on the scale of interstellar clouds, so that the formation of a star must involve the development of a near-singularity in the density distribution.  But this result was not anticipated before the numerical calculations were done by Penston and me, and also by Bodenheimer \& Sweigart at about the same time.  Even though earlier studies, notably the work of Hayashi \& Nakano in 1965, had shown a tendency for collapsing clouds to become increasingly centrally condensed, no one had anticipated the runaway development of a density singularity, and it took computers to discover this result (computers which at the time had far less computing power than your cell phone.)  So this seemingly universal feature of star formation can be regarded as a true discovery of numerical work, and as an example of how computation can discover qualitatively new phenomena.  

  A second apparently universal feature of star formation that has become clear from much computational work over the years is that, when no artificial symmetries are imposed and fully \hbox{3-dimensional} behavior is allowed to occur, we are immediately in the realm of chaotic dynamics, because only the very simplest physical systems show regular and predictable behavior.  Newton famously solved the \hbox{2-body} problem but failed to solve the \hbox{3-body} problem because it exhibits chaotic behavior, a phenomenon that is now understood largely on the basis of computational work.  Even the restricted \hbox{3-body} problem, where the third body is massless, is chaotic and can show exceedingly complex and unpredictable behavior.  Three-body interactions are almost certainly very common in star formation, and in my 1972 paper on collapse with rotation I had speculated that in reality the result might often be the formation of a triple system that decays into a binary and a single star, yielding binaries and single stars in roughly the right proportions.  Such unstable and chaotic behavior is in fact often seen in 3D simulations of the formation of systems of stars, even the first crude ones that I made in 1978, and it is not surprising because as more mass accumulates into the near-singularities or `sink particles', the system becomes increasingly like a gravitational \hbox{n-body} system whose dynamics is well known to be chaotic.  In addition to chaotic gravitational dynamics, another source of chaotic behavior that can be important in star formation is the development of fluid-dynamical turbulence in star-forming clouds.

  Because of these effects, even the simplest extension of star formation modeling from one star forming in isolation to two stars forming in a binary system involves chaotic dynamics.  Not only is the gravitational dynamics of the gas circulating around the forming stars intrinsically chaotic, but the gas flow can become turbulent, in which case there are two sources of chaotic behavior in the system.  Gravitational and MHD instabilities in the gas orbiting around the forming stars might introduce yet additional sources of chaotic behavior.  As a result, the formation of a binary system is not a deterministic or predictable process in its details -- every calculation will produce a different result.  Therefore we can only hope to predict the statistical properties of binary systems.  Large 3D simulations are beginning to be able to do this, and they have already yielded some realistic-looking results for the distributions of binary properties, including a very wide spread in separations resulting from the chaotic dynamics.  Similar considerations also apply to predicting stellar masses -- we can't predict the mass of an individual star, whose accretion history may be very chaotic and irregular, but we might be able to predict the IMF of a large ensemble of stars if we can include enough of the relevant physics.  Again, large numerical simulations are beginning to be able to address this problem.  Of course, extensive computations are needed to do these things, and powerful computers are required; computers with the power of cell-phone processors are no longer adequate.

  These examples illustrate that, in my view, the most valuable contributions that computing can make to science are not numbers but new discoveries and insights.  So I hope that the participants in this meeting who are doing computational work on star formation keep this in mind, and I look forward to learning about many new discoveries made by computational work.

\end{document}